\documentclass[prl,showpacs,superscriptaddress,twocolumn,floatfix]{revtex4}
\usepackage{graphicx,dcolumn}
\newcommand{\thgg}{\theta_{\gamma\gamma}}
\newcommand{\qcm}{q_\mathrm{cm}}
\newcommand{\qprcm}{q'_\mathrm{cm}}
\newcommand{\qprcmsq}{q'_\mathrm{cm} \! ^2}
\newcommand{\pltperp}{P_{LT}^{\perp }}
\newcommand{\pttperpprim}{P_{TT}^{' \perp}}
\newcommand{\pltperpprim}{P_{LT}^{' \perp}}
\newcommand{\pltzprim}{P_{LT}^{' z}}
\newcommand{\ptt}{P_{TT}}
\newcommand{\pttperp}{P_{TT}^{\perp}}
\newcommand{\pltz}{P_{LT}^{z}}
\newcommand{\pllptte}{P_{LL}-P_{TT}/\epsilon}
\newcommand{\pll}{P_{LL}}
\newcommand{\plt}{P_{LT}}
\newcommand{\pzcm}{P_z^\mathrm{cm}}
\newcommand{\pxcm}{P_x^\mathrm{cm}}
\newcommand{\pycm}{P_y^\mathrm{cm}}
\DeclareMathAccent{\pol}{\mathord}{letters}{"7E}
\begin{document}
\title{Measurement of the Beam-Recoil Polarization in Low-Energy Virtual Compton Scattering from the Proton}
\def\kph{\affiliation{Institut f\"ur Kernphysik, Johannes
  Gutenberg-Universit\"at Mainz, D-55099 Mainz, Germany}}
\def\gent{\affiliation{Department of Physics and Astronomy,
  University of Gent, B-9000 Gent, Belgium}}
\def\clermont{\affiliation{Clermont Universit\'e, Universit\'e Blaise Pascal, CNRS/IN2P3, LPC, BP 10448, F-63000 Clermont-Ferrand, France}}
\def\zagreb{\affiliation{Department of Physics, University of Zagreb, SI-10002
  Zagreb, Croatia}}
\def\cea{\affiliation{CEA IRFU/SPhN Saclay, F-91191 Gif-sur-Yvette
  Cedex, France}}
\def\tum{\affiliation{Physik-Department, Technische Universit\"at M\"unchen,
  D-85748 Garching, Germany}}
\def\pavia{\affiliation{Dipartimento di Fisica,
  Universit\'a degli Studi di Pavia and INFN, I-27100 Pavia,
  Italy}}
\def\stefan{\affiliation{Jo\v zef Stefan Institute, SI-1000 Ljubljana, Slovenia}}
\def\unil{\affiliation{Department of Physics, University of Ljubljana, SI-1000 Ljubljana, Slovenia}} 
\def\iss{\affiliation{Institute of Space Science, RO 76900,
  Bucharest-Magurele, Romania}}
\author{L. Doria}
\thanks{Now at TRIUMF, Vancouver, BC  V6T 2A3, Canada}\kph     
\author{P.~Janssens}
\thanks{Ph. D. fellowship Research Foundation - Flanders (FWO)}
\gent
\author{P.~Achenbach}\kph   
\author{C.~Ayerbe~Gayoso}\kph
\author{D.~Baumann}\kph    
\author{I.~Bensafa}\clermont
\author{M.~Benali}\clermont
\author{J.~Beri\v{c}i\v{c}}\stefan
\author{J.\,C.~Bernauer}\kph  
\author{R.~B\"ohm}\kph
\author{D.~Bosnar}\zagreb   
\author{L.~Correa}\clermont
\author{N.~D'Hose}\cea     
\author{X.~Defa\"y}\clermont
\author{M.~Ding}\kph      
\author{M.\,O.~Distler}\kph
\author{H.~Fonvieille}\clermont
\author{J.~Friedrich}\kph
\author{J.\,M.~Friedrich}\tum 
\author{G.~Laveissi\`ere}\clermont
\author{M.~Makek}\zagreb    
\author{J.~Marroncle}\cea
\author{H.~Merkel}\email{merkel@kph.uni-mainz.de}\kph
\author{M.~Mihovilovi\v{c}}\kph
\author{U.~M\"uller}\kph
\author{L.~Nungesser}\kph   
\author{B.~Pasquini}\pavia
\author{J.~Pochodzalla}\kph  
\author{O.~Postavaru}\kph\iss
\author{M.~Potokar}\stefan   
\author{D.~Ryckbosch}\gent
\author{S.~S\'anchez Majos}\kph
\author{B.\,S.~Schlimme}\kph
\author{M.~Seimetz}\kph    
\author{S.~\v{S}irca}\unil\stefan
\author{G.~Tamas}\kph     
\author{R.~Van~de~Vyver}\gent
\author{L.~Van~Hoorebeke}\gent 
\author{A.~Van~Overloop}\gent
\author{Th.~Walcher}\kph    
\author{M.~Weinriefer}\kph
\collaboration{A1 Collaboration}
\begin{abstract}
  Double-polarization observables in the reaction $\pol{e}p\rightarrow
  e^{\prime}\pol{p}{}^{\, \prime}\gamma$ have been measured at
  $Q^{2}=0.33\,(\mathrm{GeV}/c)^2$. The experiment was performed at
  the spectrometer setup of the A1 Collaboration using the 855\,MeV
  polarized electron beam provided by the Mainz Microtron (MAMI) and a
  recoil proton polarimeter. From the double-polarization observables
  the structure function $\pltperp$ is extracted for the first time,
  with the value $(-15.4 \, \pm 3.3 _{\mathrm{(stat.)}} \, ^{+1.5}_{-2.4} \,
  _{\mathrm{(syst.)}})\,\mathrm{GeV}^{-2}$, using the low-energy theorem for
  Virtual Compton Scattering. This structure function provides a
  hitherto unmeasured linear combination of the generalized
  polarizabilities of the proton.
\end{abstract}
\maketitle
\section{Introduction}
\label{sec-intro}

Polarizabilities parametrize the response of systems composed of
charged constituents to electric and magnetic external fields. For
the proton they contain information about the QCD interaction in the
very low momentum-transfer domain where the coupling constant
$\alpha_{\mathrm{strong}}$ diverges. Since no static field of sufficient
strength can be produced experimentally they are measured by means of
Real Compton Scattering (RCS). Now due to the availability of powerful electron accelerators also
Virtual Compton Scattering (VCS) can be investigated. VCS allows for
the determination of Generalized Polarizabilities (GPs) as function of
the initial photon virtuality $Q^2$ as first pointed out
in~\cite{Arenhoevel} for atomic nuclei and in~\cite{Guichon:1995pu}
for nucleons. Just as the form factors $G_E$ and $G_M$ give access to
the spatial density of charge and magnetization in the nucleon, the
GPs give access to such densities for a nucleon deformed by an applied
quasi-static electromagnetic field~\cite{Guichon:1995pu,
  Guichon:1998xv, Gorchtein:2009qq, Holstein:2013kia}. Out of the six
lowest-order GPs of the proton, the electric and magnetic GPs have
already been the subject of experimental investigation at
MAMI~\cite{Roche:2000ng, Janssens:2008qe},
Bates~\cite{Bourgeois:2011zz} and JLab~\cite{Fonvieille:2012cd}. The
four remaining ones, called the spin GPs, are still totally unknown
experimentally. This Letter presents the first measurement of a
double-polarization observable in VCS, with the aim of gaining insight
into the spin-GP sector of the nucleon for the first time.

\section{Formalism and Notation}

\begin{figure}
\centerline{\includegraphics[width=\columnwidth]{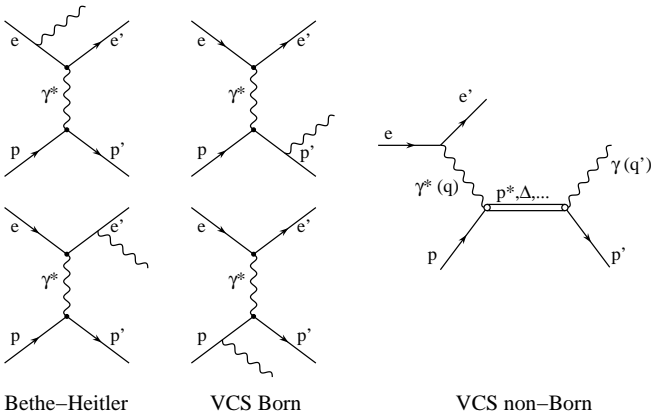}}
\caption{Contributions to the photon electroproduction amplitude. The
  VCS non-Born is parametrized by the GPs, while the BH+VCS Born
  contribution (BH+B) contains no GP effect and is entirely calculable
  in QED.}
\label{fig:1}
\end{figure}

VCS is experimentally accessed through the photon electroproduction
reaction $ep\rightarrow e'p^{\, \prime}\gamma$. At low energy it can
be decomposed into a dominant Bethe-Heitler (BH) part, a VCS Born (B)
part and a VCS non-Born (nB) part, as shown in Fig.~\ref{fig:1}. The
contributions of the Bethe-Heitler and Born processes (BH+B) can be
exactly calculated using as input only the form factors of the
nucleon. The non-Born part is parametrized at the first order in the
real photon momentum $q^{\prime}$ by six GPs. With an unpolarized
cross section measurement only two linear combinations of the GPs can
be determined. For extracting all the GPs, double-polarization
measurements are required. In this experiment, the beam-recoil
polarization asymmetries were measured in the reaction
$\pol{e}p\rightarrow e'\pol{p}^{\, \prime}\gamma$.

The main kinematical variables are defined in the $(\gamma p)$
center-of-mass (CM): the modulus of the momentum of the virtual photon
$\qcm$, of the outgoing photon $\qprcm$, and the polar angle $\thgg$
between the two photons. The virtual photon polarization $\epsilon$
and the angle $\varphi$ between the leptonic and reaction planes
complete the kinematics.

The Low Energy Theorem (LET) for the double-polarization observables
was developed in~\cite{Guichon:1998xv, Vanderhaeghen:1997bx} and is
only briefly recalled here. Experimentally, the double-polarization
observable is determined via
\begin{equation}
\mathcal{P}^\mathrm{cm}_{\hat{\imath}} = \frac{d^5\sigma(h, \hat{\imath}) -
  d^5\sigma(h,-\hat{\imath})} {d^5\sigma(h,\hat{\imath}) +
  d^5\sigma(h,-\hat{\imath})} , 
\label{eq01}
\end{equation}
where $\hat{\imath}=x,\ y,\ z$ is the CM axis for the recoil proton
polarization component, $h = \pm {1 \over 2}$ the beam helicity and
$d^5\sigma(h,\hat{\imath})$ the doubly polarized ($\pol{e}p\rightarrow
e^{\prime}\pol{p}{}^{\, \prime}\gamma$) cross section. The LET
expansion, which is valid below pion threshold, leads to:
\begin{equation}
\mathcal{P}^\mathrm{cm}_{\hat{\imath}} = \frac{\Delta d^{5}\sigma^{\mathrm{BH+B}} + \phi
  \qprcm \Delta\mathcal{M}^{\mathrm{nB}}(h,\hat{\imath}) + \mathcal{O}
( \qprcmsq )}
{2d^{5}\sigma} , 
\label{eq02}
\end{equation}
where $\Delta d^{5}\sigma^{\mathrm{BH+B}}$ is the difference of the
doubly polarized cross sections
$d^{5}\sigma^{\mathrm{BH+B}}(h,\hat{\imath}) -
d^{5}\sigma^{\mathrm{BH+B}}(h,-\hat{\imath})$ and $d^5\sigma$ is the
unpolarized $(ep \to e' p' \gamma)$ cross section. ($\phi \qprcm$) is
a phase-space factor. The non-Born terms
$\Delta\mathcal{M}^{\mathrm{nB}}$ are linear combinations of the VCS
structure functions $\pltperp$, $\pttperp$, $\pttperpprim$,
$\pltperpprim$, $\pltz$, $\pltzprim$, which can be expressed as linear
combinations of the six GPs. In particular, $\pltperp$ is a linear
combination of the structure functions $\pll$ and $\ptt$, where $\pll$
is proportional to the electric GP, and $\ptt$ is a combination of two
spin GPs: $P^{(M1,M1)1}$ and $P^{(L1,M2)1}$, the latter corresponding
to $\gamma_{E1M2}$ in the RCS limit of $Q^2\rightarrow 0$. For more
detailed formulas, we refer the reader to refs.~\cite{Guichon:1998xv,
  Vanderhaeghen:1997bx}.

\section{Experiment}

The experiment was performed at the spectrometer setup of the A1
collaboration at MAMI~\cite{Blomqvist:1998xn} and details of the
analysis can be found in~\cite{DoriaPhD:2008, JanssensPhD:2007}.
Table~\ref{tab:1} summarizes the two kinematical setups of the
experiment.

\begin{table}
  \caption{\label{tab:1} Parameters of the spectrometer setups: $p$ is the
    central momentum and $\theta$ the in-plane angle. Both settings are
    centered on the nominal kinematics defined by
    $\qcm =600\,\mathrm{MeV}/c$, $\epsilon = 0.64$,
    $\qprcm =90\,\mathrm{MeV}/c$ 
    and $\varphi=180^{\circ}$. They differ in the covered in-plane
    CM-angle $\thgg$.}
  \begin{tabular*}{\columnwidth}{@{\extracolsep{\fill}}lccccc}
    \hline\hline
    \noalign{\smallskip}
    Setup & Beam & \multicolumn{2}{c}{Spectrometer A} & \multicolumn{2}{c}{Spectrometer B}\\
    &$E$ & $p_{\mathrm{proton}}$    & $\theta_{\mathrm{proton}}$ 
    & $p_{\mathrm{electron}}$   & $\theta_{\mathrm{electron}}$  \\
  &(MeV)& $(\mathrm{MeV}/c)$ &     & $(\mathrm{MeV}/c)$ &      \\ 
  \hline
  \noalign{\smallskip}
  VCS90a & 855 & 620 & 34.1$^{\circ}$ & 546 & 50.6$^{\circ}$ \\
  VCS90b & 855 & 645 & 38.0$^{\circ}$ & 539 & 50.6$^{\circ}$ \\
  \noalign{\smallskip}
  \hline
  \hline
  \end{tabular*}
\end{table}

The polarized electron beam was delivered by MAMI with an electron
energy of $E =854.6\,\mathrm{MeV}$ and a longitudinal beam
polarization of $P_{b} = 70\%$ in average. The beam
polarization was determined by a M\o{}ller-polarimeter and was flipped
on a random basis with 1\,Hz on average to avoid false asymmetries. A
beam current of $22\,\mu\mathrm{A}$ was directed on a liquid hydrogen
target with a length of $5\,\mathrm{cm}$. The beam was rastered across
the target to avoid local boiling.

Two particles were detected in coincidence: the scattered electron in
spectrometer B with a solid angle of 5.6\,msr and a momentum
resolution of $10^{-4}$, and the recoil proton in spectrometer A with
a solid angle of 21\,msr and the same momentum resolution. Thanks to
the good timing resolution, of 0.9\,ns (FWHM) for the coincidence
time, no further particle identification was necessary. Behind the
focal plane of spectrometer A, a proton polarimeter determined the
transverse components of the proton polarization in the focal plane
(see refs.~\cite{Pospischil:2000pu, Pospischil:2002mj}).

The reaction was further identified by the missing mass squared,
\textit{i.e.} the squared mass $M_X^2$ of the missing particle $X$ in
the $(e p \to e' p' X)$ process. The $M_X^2$ distribution shows a
clean peak at zero from photon electroproduction, which is well
separated from the pion peak from $\pi^0$ electroproduction.

For the analysis, events with a missing mass squared of
$-1000\,\mathrm{MeV}^2/c^4 < M_X^2 < 4000\,\mathrm{MeV}^2/c^4$ and a
coincidence time of $-1.5\,\mathrm{ns} < t_{AB} < 1.5\,\mathrm{ns}$
were accepted as VCS events. Events from the side bands of the
coincidence time distribution were used to estimate the background
contribution due to random coincidences. A cut was also required to
eliminate the target endcaps.

For the determination of the polarization observables only events
below the pion threshold were selected, by a cut of $\qprcm <
126\,\mathrm{MeV}/c$. A standard set of further cuts were applied to
ensure a clean reconstruction within the acceptance of the recoil
polarimeter and to select the region of large analyzing power of the
polarimeter~\cite{DoriaPhD:2008}. A sample of about 77\,000 VCS
events survived the cuts.

\section{Beam-recoil Polarization Analysis} 

\label{sec-beamrecoil}

With the polarimeter, for each event the direction of the secondary
scattering process in the carbon analyzer was determined. This
direction is given by the polar and azimuthal scattering angles
$\Theta_s$ and $\Phi_s$. The distribution of events is given by:
\begin{eqnarray}
\sigma(\Theta_{s},\Phi_{s},E_{p})=\sigma_{0} [ 1 
  &\!\!\!+\!\!\!&h\,P_{b}\,A_C(\Theta_{s},E_{p})P_{y}^{\mathrm{fp}}\cos\Phi_{s} 
\nonumber\\ 
  &\!\!\!-\!\!\!&h\,P_{b}\,A_C(\Theta_{s},E_{p})P_{x}^{\mathrm{fp}}\sin\Phi_{s} ],~~~
\label{fppphi}
\end{eqnarray}
it depends on the known analyzing power of the carbon analyzer
$A_{C}(\Theta_{s},E_{p})$ (see \cite{DoriaPhD:2008,Pospischil:2000pu})
and the transverse components of the proton double polarization
observable in the focal plane, $P_{y}^{\mathrm{fp}}$ and
$P_{x}^{\mathrm{fp}}$. For a given set of CM polarizations
$P_{x,y,z}^\mathrm{cm}$, the focal plane transverse components
$P_{x}^{\mathrm{fp}}$ and $P_{y}^{\mathrm{fp}}$ can be calculated by
Lorentz transformation, rotation and ray-tracing of the spin
precession in the magnetic field of the spectrometer. Thus, the CM
polarizations can be fitted to the distribution of the azimuthal angle
$\Phi_{s}$ by a standard maximum likelihood method. This is the first
step of the analysis.

In principle, the statistical ensemble contains the information for
all three CM components of the polarization, since events with
different orientation of the scattering plane have different paths in
the magnetic field of the spectrometer, resulting in different
transverse components in the focal plane. A detailed simulation
showed, however, that the longitudinal component $\pzcm$ cannot be
reconstructed with sufficient resolution. Therefore this component was
fixed in the analysis for each event to the value given by the BH+B
calculation, \textit{i.e.} $\pzcm = \Delta d^{5}\sigma^{\mathrm{BH+B}}
/2d^{5}\sigma^{\mathrm{BH+B}}$. The simulation showed that this choice
was sufficient to provide a non-biased fit of $\pycm$ and $\pxcm$. A
more realistic choice, \textit{i.e.} adding a GP effect in the
constraint on $\pzcm$, was considered only to evaluate systematic
errors.

The maximum likelihood fit yields the CM polarization components
$\pxcm$ and $\pycm$. The fit is made separately in five $\thgg$ bins
to have sufficient statistical significance per bin.

%
\begin{table}
\begin{center}
  \caption{\label{tab:2} Results for the double-polarization
    observables. $\pxcm$(raw) and $\pxcm$(proj.) are the fitted
    $\pxcm$ component before and after the projection to the nominal
    kinematics, respectively. $\Delta \pxcm$ (stat.) is the
    statistical error on $\pxcm$(proj.), while $\Delta \pxcm$ (syst.)
    are systematic errors (see text). Negative $\thgg$ values are
    conventional for $\varphi=180^{\circ}$.}
\begin{tabular*}{\columnwidth}{@{\extracolsep{\fill}}rlrrrrr}
\hline
\hline
\multicolumn{2}{c}{$\thgg$} & $-170^{\circ}$ & $-150^{\circ}$ & $-130^{\circ}$ & $-110^{\circ}$ & $-90^{\circ}$\\ 
\hline 
$\pycm$&(raw)              & $0.047$    & $0.012$       & $-0.043$       & $0.020$       & $-0.020$    \\
$\Delta \pycm$ &(stat.) & $\pm 0.066$    & $\pm 0.053$    & $\pm 0.038$    & $\pm 0.041$    & $\pm 0.050$    \\
\hline
$\pxcm$&(raw)       & $-0.220$    & $-0.269$    & $-0.215$    & $-0.177$    & $-0.067$    \\
$\pxcm$&(proj.)     & $-0.209$    & $-0.257$    & $-0.201$    & $-0.142$    & $-0.041$    \\  
$\Delta \pxcm$ &(stat.)  & $\pm 0.049$ & $\pm 0.040$ & $\pm 0.030$ & $\pm 0.027$ & $\pm 0.027$ \\   
$\Delta \pxcm$ &(syst.1) & $\pm 0.001$ & $\pm 0.011$ & $\pm 0.007$ & $\pm 0.009$ & $\pm 0.004$ \\  
$\Delta \pxcm$ &(syst.2)& $\pm 0.030$ & $\pm 0.001$ & $\pm 0.003$ & $\pm 0.020$ & $\pm 0.030$ \\ 
$\Delta \pxcm$ &(syst.3) & $\pm 0.010$ & $\pm 0.020$ & $\pm 0.020$ & $\pm 0.020$ & $\pm 0.010$ \\ 
\hline
\hline
\end{tabular*}
\end{center}
\end{table}

Table~\ref{tab:2} summarizes the results. The obtained values for
$\pycm$ are compatible with zero within the uncertainties; this is
consistent with the requirement that $\pycm$ has to vanish in strict
in-plane kinematics. Globally, $\pycm$ has a negligible sensitivity to
the GPs and almost all the new information is carried by $\pxcm$,
through the term $\Delta\mathcal{M}^{\mathrm{nB}}(h,\hat{x})$ which is
of the form:
\begin{eqnarray}
\Delta\mathcal{M}^{\mathrm{nB}}(h,\hat{x}) &=& h \left( a_{1}^{x} \pltperp +
a_{2}^{x} \pttperp \right.\nonumber\\
&& ~\left. + a_{3}^{x} \pttperpprim +
a_{4}^{x} \pltperpprim \right)         \! , 
\label{dpo_x}
\end{eqnarray}
with $a_i^x$ being known kinematical
coefficients~\cite{Vanderhaeghen:1997bx}.

Figure~\ref{fig:2} displays the measured $\pxcm$ component as five
solid points. These points have been projected to the nominal
kinematics ($\qcm, \epsilon, \qprcm,$ and $\varphi$ of
Table~\ref{tab:1}), completed by the values of $\thgg$ of
Table~\ref{tab:2}. This projection is based on the expected LET
behavior of the polarization observables as a function of the
kinematics.

The statistical error is provided by the fit. Systematic errors on
$\pxcm$ have been determined as coming from: 1) a beam polarization
uncertainty of $\pm$ 1.2\% ; 2) changing the constraint on $\pzcm$; 3)
uncertainties in the kinematical projection. Other systematic
effects, due to instrumental asymmetries in the proton polarimeter or
due to random coincidences under the time peak were found to be
negligible.

\section{Structure function Analysis}
\label{sec-structurefunc}

As a next step, a fit was performed with the aim of determining
individual GPs (including the spin GPs). The principle is again to use
the likelihood method, this time fully unbinned. The non-Born terms
$\Delta\mathcal{M}^{\mathrm{nB}}(h,\hat{x})$ and
$\Delta\mathcal{M}^{\mathrm{nB}}(h,\hat{y})$ in the numerator of
$\pxcm$ and $\pycm$ were replaced by their analytical expression in
terms of the GPs~\cite{Vanderhaeghen:1997bx}. The cross section
$d^5\sigma$ in the denominator of $\pxcm$ and $\pycm$ was fixed to its
value given by the unpolarized LET expression, using our previously
measured structure functions ($\pllptte$) and
$\plt$~\cite{Janssens:2008qe}. As an outcome, it turned out that the
data were not precise enough to extract individual GPs. However, if
one uses structure functions, \textit{i.e.} combinations of GPs,
instead of GPs directly, one gets a significant result for $\pltperp$
as we show in the following final step of the analysis.

%
%
%
\begingroup
\renewcommand{\arraystretch}{1.4}
\begin{table*}
  \begin{center}
    \caption{\label{tab:3} The complete set of VCS structure functions
      at $Q^2=0.33\,(\mathrm{GeV}/c)^2$ as calculated by two models:
      (I)= DR formalism~\cite{Pasquini:2001} (with
      $\Lambda_{\alpha}=1.80\,\mathrm{GeV}$,
      $\Lambda_{\beta}=0.75\,\mathrm{GeV}$ and the MAID03 version),
      (II) = HBChPT at $\mathcal{O}(p^3)$~\cite{Hemmert:1999pz},
      including the $\pi^0$-pole term (or anomaly). Only six of these
      nine structure functions are independent, \textit{e.g.} the ones
      in the first six columns. $\pttperp$, $\pttperpprim$, and
      $\pltperpprim$ are fixed to the values of this Table when
      fitting $\pltperp$ (see text).  All calculations are done with
      the proton form factors of ref.~\cite{Mergell:1995bf}. This also
      holds for Table~\ref{tab:5}. }
    \begin{tabular}{llccccccccc}
      \hline
      \hline
      & \multicolumn{9}{c}{Model Structure Functions ($\mathrm{GeV}^{-2}$)} \\
      \hline
      \multicolumn{2}{l}{model} &  \ $\pttperp$ & \  $\pttperpprim$  & \  $\pltperp$  & \ $\pltperpprim$   & \ $\plt$  & \   $\pll$  & \ $\ptt$ & \ $\pltz$ & \ $\pltzprim$ \\
      \hline
      (I) &DR model & $0.97$ & $-0.44$ & $-10.83$ & $-1.43$   & $-2.43$ & $22.40$  &  $-1.58$ & $-1.34$ & $-1.21$ \\
      (II) &HBChPT $\mathcal{O}(p^3)$   &  $2.05$ &  $\ \ 0.62$  &  $-10.57$ & $-4.21$   & $-5.34$ & $15.07$  &  $-6.89$  &  $-3.03$   & $-0.86$  \\
      \hline
      \hline
    \end{tabular}
  \end{center}
\end{table*}
\endgroup

The unbinned maximum likelihood method is again used.
The non-Born terms
$\Delta\mathcal{M}^{\mathrm{nB}}(h,\hat{x})$ and 
$\Delta\mathcal{M}^{\mathrm{nB}}(h,\hat{y})$  
are replaced by their analytical expressions in terms of the 
structure functions, as \textit{e.g.} in Eq.~(\ref{dpo_x}). 
The denominators of  $\pxcm$ and $\pycm$ are treated as above. 

Exploratory fits showed that $\Delta\mathcal{M}^{\mathrm{nB}}$ 
is sensitive mainly to $\pltperp$, among the four structure 
functions entering  Eq.~(\ref{dpo_x}). Therefore, the other three: 
$\pttperp, \pttperpprim$ and $\pltperpprim$  
cannot be fitted. However, their influence can be investigated by
inserting several model predictions and fitting $\pltperp$
only. This implies a model dependence of the extracted results, but 
we show in the following that it is relatively small compared 
to the statistical uncertainty.

The structure functions that need to be fixed are only $\pttperp,
\pttperpprim$ and $\pltperpprim$, \textit{i.e.} the ones appearing in
$\Delta\mathcal{M}^{\mathrm{nB}}(h,\hat{x})$ and
$\Delta\mathcal{M}^{\mathrm{nB}}(h,\hat{y})$, except $\pltperp$. The
maximum likelihood fit was done with three rather different
assumptions for these fixed structure functions. In fit ``I'' they
were set to values calculated by the Dispersion Relation (DR)
model~\cite{Pasquini:2001}, cf. the first line of
Table~\ref{tab:3}. In fit ``II'' they were set to values calculated by
Heavy Baryon Chiral Perturbation Theory
(HBChPT)~\cite{Hemmert:1999pz}, cf. the second line of
Table~\ref{tab:3}. In fit ``III'' they were all set to zero. These
different choices lead to the following results: $\pltperp = -15.4,
-17.7$ and $-14.1\,\mathrm{GeV}^{-2}$ for fits ``I'', ``II'' and
``III'' respectively. We consider fit ``I'' as the central one,
yielding our final result for $\pltperp$, and the two other results
are used to estimate the model-dependent error.

 
The statistical error on $\pltperp$ is provided by the maximum
likelihood fit. The systematic error comes from several main sources,
which are estimated in Table~\ref{tab:4}. The first contribution is
obtained by changing the beam polarization by $\pm$ 1.2\% in the
analysis. The second contribution is estimated by performing the fit
with several form factor parametrizations~\cite{Arrington:2007ux,
  Friedrich:2003iz, Belushkin:2006qa, Bernauer:2013tpr}; the maximal
spread of the results gives the magnitude of the error, which remains
small. The third contribution is related to the treatment of $\pzcm$;
the error is obtained as the difference in the fitted result when we
fix $\pzcm$ to its BH+B value, or when a GP effect is added to it.
The fourth contribution is due to model dependence; it is determined
from the differences between the various fits (``II'' $-$ ''I'' and
``III'' $-$ ''I''). In Table~\ref{tab:4} each partial systematic error
has been symmetrized except the fourth one which is the largest and
most asymmetric. The total systematic error is calculated as the
quadratic sum of the errors of Table~\ref{tab:4} for each sign
separately.


\begin{table}
\begin{center}
\caption{\label{tab:4} Systematic errors in the extraction of $\pltperp$.}
\begin{tabular*}{\columnwidth}{@{\extracolsep{\fill}}lc}
\hline
\hline
 Error Type      &    Error Value ($\mathrm{GeV}^{-2}$)\\
\hline
 beam polarization ($\pm 1.2$\%)   &  $\mp$  0.53\\
 proton form factors     &  $\pm$  0.10 \\
 constraint on $\pzcm$       &  $\pm$  0.47 \\
fixed structure functions  & $+1.26/-2.29$ \\
\hline
 Total systematic error &  $+1.45/-2.40$ \\
\hline
\hline
\end{tabular*}
\end{center}
\vspace{-2mm}
\end{table}


Our final result for $\pltperp$ is presented in Table~\ref{tab:5}. It
is compared to theoretical values from HBChPT and DR calculations.
The absolute value of the result is larger than in most theoretical
calculations. Some features of the models are worth noting: In HBChPT
some of the GPs have a bad convergence with respect to the order of
the calculation~\cite{Kao:2002, Kao:2004us}, and this may affect the
model value of $\pltperp$. In the DR model the spin GPs are entirely
fixed, but the scalar GPs contain an unconstrained part that has to be
fitted from experiment. In particular $\pltperp$ depends, via the
structure function $\pll$, on the free parameter $\Lambda_{\alpha}$
which determines the electric GP. Table~\ref{tab:5} shows this
dependence for a realistic range of values for $\Lambda_{\alpha}$. We
note that the DR model has a lower limit for $\pltperp$ of $-13.1\,\mathrm{GeV}^{-2}$ (for $\Lambda_{\alpha}= \infty$).

\begin{figure}
\begin{center}
\resizebox{\columnwidth}{!}{\includegraphics{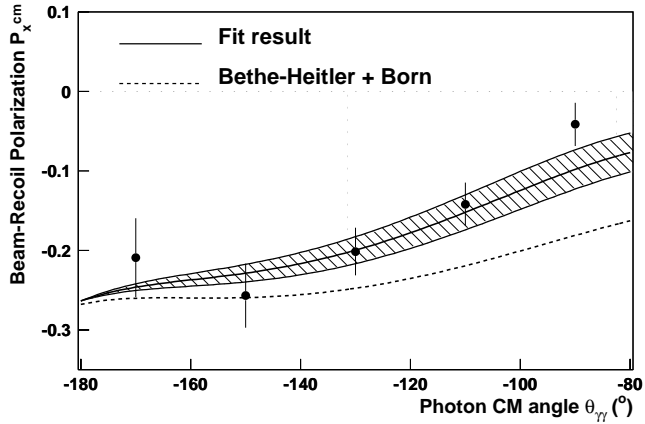}}

\end{center}
\vspace{-3mm}   
\caption{Measured recoil proton polarization component $\pxcm$ in the
  CM frame. The five points with their statistical error are the
  result of the first-step fit. The solid curve is calculated using
  our result for$\pltperp$ (see text); the shaded band represents the
  statistical uncertainty. The dashed curve is the BH+B calculation of
  $\pxcm$, \textit{i.e.} without any GP effect. }
\label{fig:2}
\end{figure}

A graphical representation of our result is shown in Fig.~\ref{fig:2}.
The central solid curve is obtained by calculating the polarization
component $\pxcm$ at the nominal kinematics, based on
Eqs.~(\ref{eq02}) and (\ref{dpo_x}).  The calculation uses the results
of fit ``I'' (see above), \textit{i.e.}  $\pltperp
=-15.4\,\mathrm{GeV}^{-2}$ and the other three structure functions set
to their DR value of Table~\ref{tab:3}. Using the results of fit
``II'' instead of ``I'' yields a very similar curve.  The deviation
from the BH+B calculation (dashed curve) is a clear signature of the
polarizability effect.

%
\begingroup
\renewcommand{\arraystretch}{1.4}
\begin{table}
  \begin{center}
    \caption{\label{tab:5} Our measured value of $\pltperp$ and
      several model predictions at
      $Q^{2}=0.33\,(\mathrm{GeV}/c)^2$. For the DR model, the (a),
      (b), (c) cases correspond to different values of the
      $\Lambda_{\alpha}$ parameter: $0.6$, $1.2$, and
      $1.8\,\mathrm{GeV}$ respectively (see text). }
\begin{tabular*}{\columnwidth}{@{\extracolsep{\fill}}ll}
      \hline
      \hline
\ &  $\pltperp$ ($\mathrm{GeV}^{-2}$) \\
      \hline
This experiment \ \ \ \ \ \ \ \    & $-15.4 \, \pm  3.3 _{\mathrm{(stat.)}} \, ^{+1.5}_{-2.4} \, _{\mathrm{(syst.)}}$ \\
\hline
DR model~\cite{Pasquini:2001}  &   $-3.7$ (a) \  , \  $-8.7$ (b) \ , \ $-10.8$ (c)   \\
HBChPT $\mathcal{O}(p^3)$~\cite{Hemmert:1999pz}  &  $-10.6$ \\
      \hline
      \hline
    \end{tabular*}
  \end{center}
\end{table}
\endgroup


In conclusion, we have measured for the first time double-polarization
observables in VCS from the proton below the pion threshold. The
analysis was based on the theoretical formulation of the LET for
polarized VCS, and the experimental use of recoil proton polarimetry.
A clear polarizability effect was observed in the $\pxcm$ polarization
component. We extracted one new structure function, $\pltperp$, and
found a value that is larger in magnitude than most theoretical
calculations. Therefore, this measurement provides a valuable and
entirely new constraint for models of nucleon structure; although it
does not allow one to further disentangle the scalar and spin GPs of
the proton.

\vskip 2 mm

We acknowledge the MAMI accelerator group for the outstanding support.
This work was supported in part by the FWO-Flanders (Belgium), the
BOF-Gent University, the Deutsche Forschungsgemeinschaft with the
Collaborative Research Center 1044, the Federal State of
Rhineland-Palatinate and the French CEA and CNRS/IN2P3.

\bibliographystyle{apsrev.bst} \bibliography{vcspol}
\end{document}